\documentclass[aps%
 ,twocolumn%
 ,secnumarabic%
,amssymb, amsmath,nobibnotes, aps, pre, floatfix,superscriptaddress]{revtex4-1}

\usepackage{graphicx,epsfig}
\usepackage{amsmath}
\usepackage{amsfonts}
\usepackage{fancyhdr}
\usepackage{hyperref}
\usepackage{epstopdf}

\begin{document}

\pagestyle{fancy}
\lhead{}
\rhead{V.Navas and E.Vives}
\lfoot{}
\rfoot{}
\title{Influence of the aspect ratio and boundary conditions on universal finite size scaling functions in the athermal metastable 2d Random Field Ising Model}
\author{V\'ictor Navas Portella} 
\affiliation{Departament d'Estructura i Constituents de la Mat\`eria, Facultat de F\'{\i}sica, Universitat de Barcelona, Diagonal
645, 08028 Barcelona, Catalonia, Spain.}
\affiliation{Centre de Recerca Matem\`atica, Edifici C, Campus Bellaterra, E-08193 Bellaterra, Catalonia, Spain.}
\author{Eduard Vives}
\affiliation{Departament d'Estructura i Constituents de la Mat\`eria, Facultat de F\'{\i}sica, Universitat de Barcelona, Diagonal
645, 08028 Barcelona, Catalonia, Spain.}
\begin{abstract}
This work  studies universal finite size scaling functions for the number of 1d spanning avalanches in a two-dimensional disordered system with boundary conditions of different nature and different aspect ratios. For this purpose, we consider the 2d Random Field Ising Model at $T=0$ driven by the external field $H$ with athermal dynamics implemented with periodic and forced boundary conditions. We choose a convenient scaling variable $z$ that accounts for the deformation of the distance to the critical point caused by the aspect ratio. Moreover, assuming that the dependence of the finite size scaling functions on the aspect ratio can be accounted by  an additional multiplicative factor, we have been able to collapse data for different system sizes, different aspect ratios and different nature of the boundary conditions into a single scaling function $\hat{\mathcal{Q}}$.

\end{abstract}

\maketitle

\section{Introduction}
The theory of universal finite size scaling (FSS) functions has been broadly used in simulations of finite systems close to criticality in order to extrapolate numerical results to the thermodynamic limit \cite{privman1}.  In two-dimensional square lattices with rectangular shape of sizes $L_{x} \times L_{y}$, it is known that those functions depend on the aspect ratio $a=L_{y}/L_{x}$ \cite{privman4}. The propierties of universal FSS functions when the aspect ratio, as well as boundary conditions, are changed have been studied in different situations \cite{langlands}-\cite{hucht}.

In the context of percolation theory \cite{percolation}, the analysis of the effects of rectangular boundaries in the spanning probability was studied by Langlands \cite{langlands}, who showed numerically that, at the critical threshold, there exists a universal scaling function of the aspect ratio. The study was based on symmetry arguments for the crossing probabilities. Inspired by the numerical results of Langlands, Cardy \cite{cardy} derived an analytical expression using the conformal invariance of the spanning probability in 2d percolation. Other numerical studies deal with this problem and some variations in geometries and boundary conditions\cite{masihi},\cite{watanabe},\cite{sticks}. Okabe \textit{et al.} studied shape effects and boundary conditions for Ising models \cite{okabe}\cite{okabe3d} by analysing the Binder parameter and magnetization curves. A. Hucht also studied the symmetries of universal FSS functions in anisotropic systems by checking a symmetry hypothesis through Monte Carlo simulations of the two-dimensional in-plane Ising model\cite{hucht}.

To our knowledge, the effects of rectangular shape as well as different boundary conditions in the FSS functions have not been studied in the two-dimensional Random Field Ising Model (RFIM) with athermal dynamics \cite{sethna93}, where we expect that quenched disorder as well as the metastable character of the dynamics strongly determine the spanning cluster \cite{percolation}.

The athermal (T=0) RFIM has been commonly used to explain the Barkhausen noise in ferromagnetic materials\cite{reviewscienceSethna}. In many cases, ferromagnetic coupling is so strong that it is not necessary to consider temperature as a relevant parameter in the model. During the field driven magnetization process, free energy barriers are so large that thermally activated events are negligible. By considering interaction between spins, quenched disorder in the sample and external magnetic field, the RFIM offers a good explanation of hysteresis and crackling noise (avalanche dynamics).

The model is also applicable to other physical systems such as structural transitions, capillary condensation of gases in porous solids, etc \cite{reviewscienceSethna}\cite{rosVives}. All these phenomena share in common the existence of a first order phase transition and hysteresis. Metastable states separated by high energy barriers appear in the free energy landscape. 

When the external force is increased  (for example the magnetic field),  the conjugated variable (magnetization) responds discontinuously.  These non-equilibrium collective events, known as 'avalanches', are essentially due to the system jumps from one metastable state to another. The properties of these avalanches depend strongly on the quenched disorder present in the system, determining which sites of the system are more favourable to nucleate or not.

The lower critical dimension of the RFIM is known to be $d_{c}=2$ in the equilibrium case \cite{2d}. This implies that ferromagnetic order will not occur for $d\leq 2$. It is not completely settled if the lower critical dimension is still $d_{c}=2$ for the RFIM with local metastable dynamics at T=0. J.P.Sethna and coworkers opened the door to a possible existence of long range order in this case \cite{condmat96}. Nevertheless, it has been necessary to examine larger system sizes in order to determine whether the critical value of the model parameter determining the amount of the quenched disorder $\sigma_{c}$ was finite or not, in the thermodynamic limit. Recently, D.Spasojevi\'c and coworkers found numerical evidence of a critical point, below which $\left( \sigma <\sigma_{c}\right)$ the system orders ferromagnetically. These evidences were based on the FSS collapse of the curves corresponding to magnetization\cite{spasojevicPRL2011}, distribution of avalanche sizes\cite{spasojevicPRE2011} and number of spanning avalanches\cite{spasojevicPRE2014}.

In order to study the effect of the aspect ratio and boundary conditions in universal FSS functions we will focus on the number of spanning avalanches in one direction \citep{spasojevicPRE2014}\cite{vives2004}. Spanning avalanches, strictly speaking, are well defined objects only in finite size simulations with square or rectangular boundaries. They correspond to magnetization events of size $S$ that extend, at least, from one side of the system to the opposite one. They can be classified as 1d-spanning or 2d-spanning, depending on whether they span the system in one, or two spatial directions. Nevertheless in the thermodynamic limit such avalanches correspond to infinite objects not necessarily massive (i.e. $S/L^{2}\rightarrow 0$) similar to the infinite cluster \cite{percolation} within percolation theory. 

Different strategies may help in understanding the critical nature of such objects. A first possibility is to study its behaviour when changing the shape of the finite system, by simulating not only square, but rectangular boundary conditions with a certain aspect ratio $a$. A second strategy is to study spanning objects when changing the physical nature of the finite boundaries. Numerical simulations, often use Periodic Boundary Conditions (PBC). Such spatial boundaries are the most adequate to minimize finite size effects when studying  a first order phase transition that occurs due to a local nucleation process. Nevertheless, some forced boundary conditions may help in the understanding of models with propagating front dynamics. Typically this is done by keeping periodic boundaries in one direction and fixing the boundaries in the other direction to be in the two different coexisting phases. We will refer to these fixed boundary conditions as FBC.
Sepp\"al\"a et al.\cite{sepala} found evidence of front roughening effects in the equilibrium RFIM at $T=0$ with FBC without external field. The model with metastable dynamics but restricted to nucleation close to the interface, was studied in the context of depinning transition \cite{dahmen98}, and for the study of morphological changes in the invading front \cite{jiRobbins91}. The ground state of the equilibrium 3d-RFIM was studied by Middelton and Fisher using a variety of fixed boundary conditions and comparing with periodic boundaries \cite{Middleton2002}

In this work, we present a numerical study of the number of 1d-spanning avalanches  $N_{1}$ in the metastable two-dimensional RFIM using the two strategies described above. The subscript $1$ is not relevant for this paper but will be kept with consistency with previous works \cite{spasojevicPRE2014}\cite{vives2004}. The model and the simulation details are presented in Section \ref{Model}. Results corresponding to the FSS analysis for rectangular PBC are presented in Section \ref{PBC}. We will obtain FSS functions that include the dependence with the aspect ratio $a$. The analysis of FBC and different  aspect ratios is presented in Section \ref{FBC}. In this case we obtain FSS functions with two additive contributions: the first is the same as for PBC and the second one is related to a roughening transition of the interface. Finally a summary and conclusions are presented in Section \ref{Conclusions}.

\section{Model}
\label{Model}
Ising models consist of an ensemble of N interactive spins situated at the nodes of a d-dimensional lattice. Spins can take the values $ s_{i}=\pm1$. The RFIM is a variant of these kind of models which includes quenched disorder (impurities, dislocations, vacancies, etc.) that distorts the free-energy landscape. 
The Hamiltonian describing this model is:
\begin{equation}
\mathcal{H} = -\sum_{\langle ij \rangle} J s_{i}s_{j} - \sum_{i}(H+h_{i})s_{i}.
\end{equation}
The first term accounts for ferromagnetic interaction with nearest neighbours.
For simplicity, we consider $J=1$. $H$ is the external field and $\lbrace h_{i} \rbrace$ are local quenched random fields, independent and Gaussian distributed according to:
\begin{equation}
\rho(h)= \frac{1}{\sqrt{2\pi \sigma^{2}}} exp\left(-\frac{h^{2}}{2\sigma^{2}}\right),
\end{equation}
where $\sigma$ characterizes the amount of disorder present in the system. We study a 2d square lattice with rectangular shape ($N=L_{x}\times L_{y}, a=L_{y}/L_{x}$). In order to generate a metastable dynamics, it is necessary to stablish a criterion to determine under which conditions the system remains in a local minimum of the energy landscape. 
Following Sethna's rule\cite{sethna93}, a spin is stable when it is aligned with its effective field:
\begin{equation}
h_{i}^{eff}=\sum_{k}^{z} s_{k}+H+h_{i},
\label{eq:dynamics}
\end{equation} 
where the sum extends over the $z$ neighbours of the spin $s_{i}$, which in the case of a 2-d square lattice is $z=4$. Following this deterministic rule, a spin flips when its effective field changes sign. A flipping event changes the value $h_{k}^{eff}$ of its nearest neighbours and any of these could become unstable. The process continues in the same way analysing successive shells and originating an avalanche of flipping spins. The size $S$ of the avalanche corresponds to the number of spins that have changed their state whereas the duration $T$ corresponds to the number of shells needed to complete the avalanche. Simulations start with all the spins pointing down ($\lbrace s_{i}=-1\rbrace$) and $H=-\infty$. The external field is increased until a spin triggers an avalanche. Then the external field is kept constant until there are no more unstable spins and the system reaches a metastable state again. The dynamical process finishes when all the spins are pointing up ($\lbrace s_{i} =+1\rbrace $) for a very large postive value of the field $H$. Sorted list algorithm has been implemented\cite{kuntzSethna}. Since our intention is not to discuss the behaviour in the thermodynamic limit but to understand FSS relations, it has been preferable to study characteristic sizes of $L_{x},L_{y}\leq 1024$ and perform a number of averages over many ($10^{5}$) randomness realizations. The range of aspect ratios $a$ for which results presented in this work are valid ranges from $a=1/2$ to $a=2$. For illustrative purposes we will also show numerical results for smaller values of $a$ ($a=1/4$ and $a=1/8$).

\begin{figure}[h!]
\resizebox{\columnwidth}{!}{\includegraphics{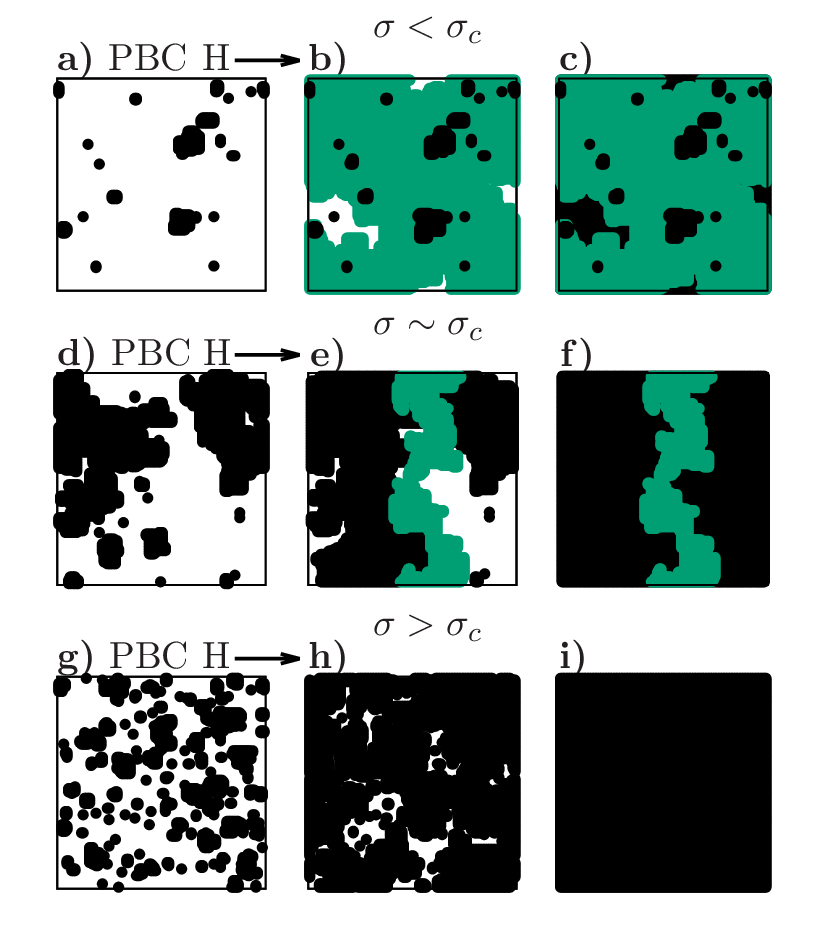}}
\caption{\label{fig:1}Sequence of configurations during the magnetization process in different disorder regimes for the RFIM with metastable dynamics for a system ($L=128$) with PBC. The values of the disorder are $\sigma=0.80,0.95,1.20$ for sequences (a-b-c),(d-e-f) and (g-h-i) respectively. External field $H$ is increased from left to right bringing the system from negative magnetization (white regions) to positive magnetization (coloured regions). Black coloured spins correspond to regions that have been transformed due to non spanning avalanches. Green regions correspond to spanning avalanches.}
\end{figure}
\section{Results}
\subsection{Periodic Boundary Conditions (PBC)}
\label{PBC}
\begin{figure}[h!]
\centering
\resizebox{\columnwidth}{!}{\includegraphics{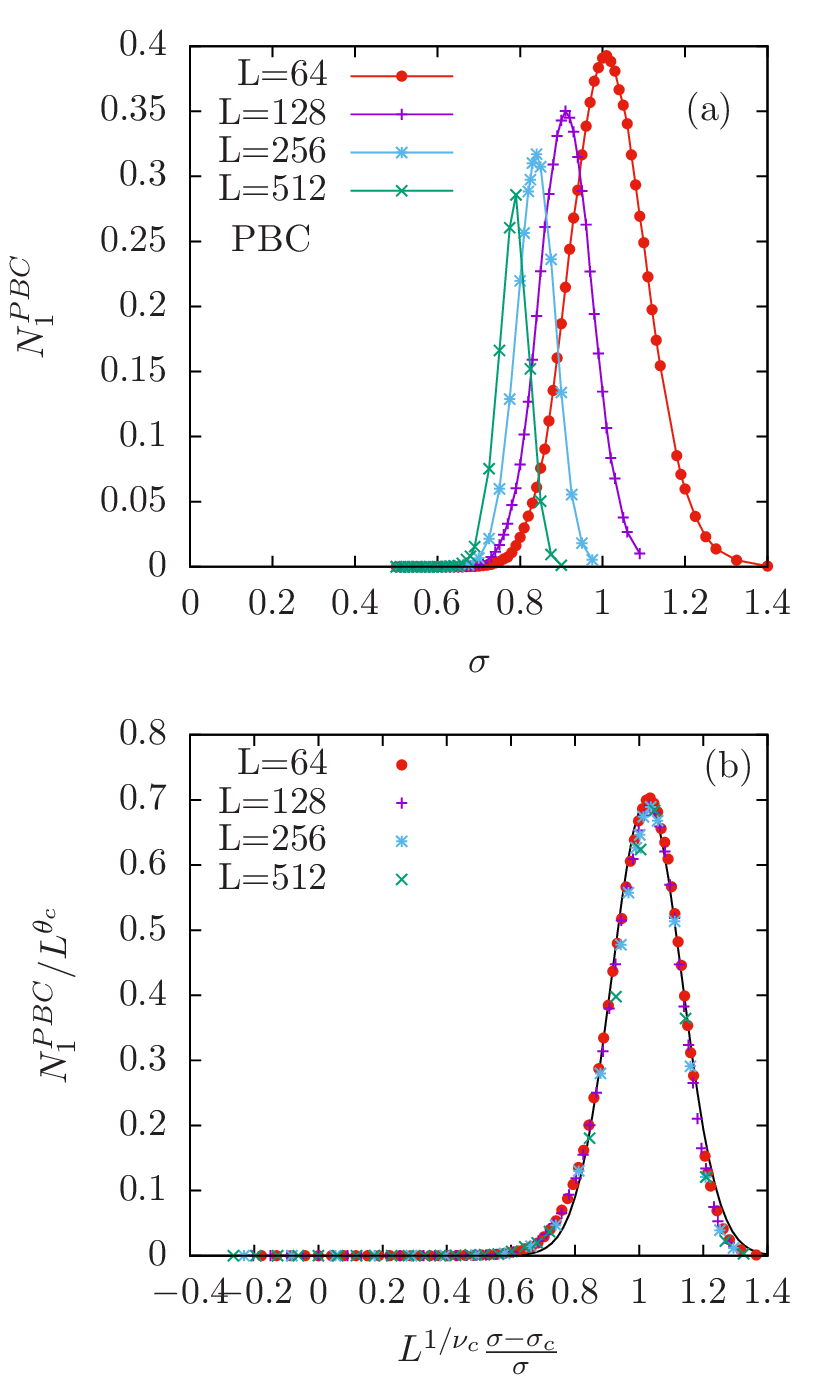}}
\caption{\label{fig:3}Panel (a) shows the average number of 1d spanning avalanches per magnetization process against the disorder $\sigma$ for different sizes in a square systems $a=1$  with PBC. Lines are guides to the eye. Panel (b) shows curve collapse for different system sizes by using Eq.~(\ref{eq:spasoscaling}). Solid black line represents a Gaussian fit of the scaled data.}
\end{figure}

The field driven, athermal RFIM with the standard PBC and $L_{x}=L_{y}\equiv L$ has been broadly studied \cite{condmat96}-\cite{spasojevicPRE2014}. By numerical analysis it is shown that there exists a continuous transition for a finite value of disorder $\sigma_{c}=0.54$ which separates two different regimes \cite{condmat96}\cite{spasojevicPRL2011}. For $\sigma<\sigma_{c}$, there is an infinite avalanche, which is usually 2-d spanning (See Fig.~\ref{fig:1}, sequence a-b-c that corresponds to increasing field values). When disorder aproaches its critical value $\sigma_{c}$, there is a peak in the average number of 1d spanning avalanches\cite{spasojevicPRE2014} (See Fig.~\ref{fig:1}, sequence d-e-f). Power-law distributions are found in some magnitudes related to avalanches in this situation (sizes and durations). For the regime $\sigma >\sigma_{c}$, the magnetization process takes place by nucleation of small domains that grow and coallesce and there is no presence of spanning avalanches (See Fig.~\ref{fig:1}, sequence g-h-i). 
The average number of 1d spanning avalanches during the full magnetization process from $H=-\infty$ to $H=+\infty$ as a function of the disorder $\sigma$ is presented in Fig.~\ref{fig:3} for square systems ($L_{x}=L_{y}=L;a=1$).

In Fig.~\ref{fig:3} (a), the number of 1d spanning avalanches exhibits a peak at the value of disorder $\sigma_{c}(L)$ as found by Spasojevi\'c et al. \cite{spasojevicPRE2014}.
These peaks present the following scaling behaviour:
\begin{equation}
N_{1}^{PBC}(\sigma,L) = L^{\theta_{c}}\mathcal{P}\left( L^{1/\nu_{c}} \frac{\sigma-\sigma_{c}}{\sigma} \right),
\label{eq:spasoscaling}
\end{equation}
where $\sigma_{c}=0.54 \pm 0.03$, $\frac{1}{\nu_{c}}=0.19 \pm 0.02$, $\theta_{c}=-0.10 \pm 0.03$ and $\mathcal{P}$ is the FSS function related to the number of 1d spanning avalanches for PBC (See Fig.~\ref{fig:3}(b))). These values are compatible with those of \cite{spasojevicPRE2014}.The error bars of the fitted parameters account for variations that still offer acceptable collapses. We use the calligraphic letter $\mathcal{P}$ for the scaling functions with PBC. Note that in this case, contrarily to what happens in the 3d case \citep{vives2004}, the number of 1d spanning avalanches vanishes in the thermodynamic limit due to the negative sign of the exponent $\theta_{c}$.

\begin{figure}[h!]
\resizebox{\columnwidth}{!}{\includegraphics{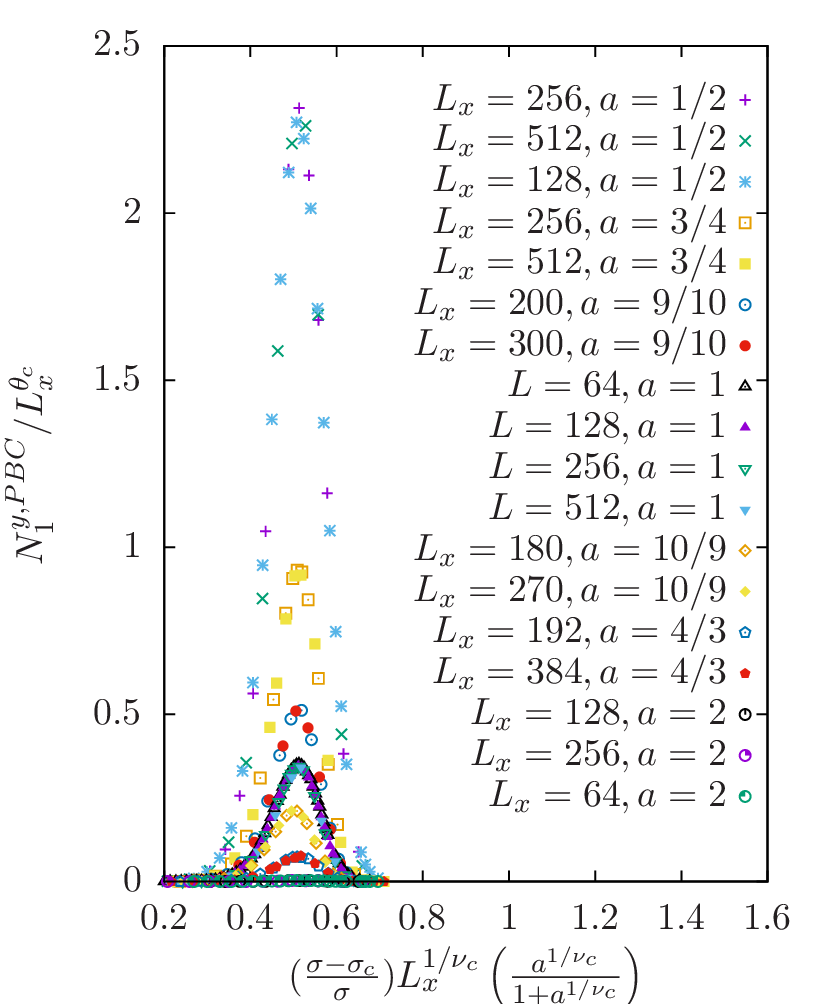}}
\caption{\label{fig:escalatprevi}Curve collapses $\mathcal{P}_{a}^{y}(z)$ for the number of 1d spanning avalanches in the $\hat{y}$ direction with the same aspect ratio $a$ and different sizes.}
\end{figure}

In the present work, we will study PBC with rectangular shape and thus we have to separate  spanning avalanches in the $\hat{y}$ and $\hat{x}$ direction. Therefore we will measure $N_{1}^{y;PBC}(\sigma,L_{x},a)$, $N_{1}^{x;PBC}(\sigma,L_{x},a)$, where we have chosen a dependence on the disorder $\sigma$, the horizontal width $L_{x}$ and the aspect ratio $a=L_{y}/L_{x}$. In order to perform the FSS analysis we need to find a scaling variable for this rectangular case. The geometric average length $L=\sqrt{L_{x}L_{y}}$ is used in \cite{sticks} for FSS of asymmetric systems of percolating sticks. We have checked that, with this choice of the scaling length it is not possible to achieve collapses for different aspect ratios. We propose an alternative choice of the scaling length that enables to collapse curves for different sizes as well as aspect ratios $a$. 
Due to the asymmetry ($a\neq 1$), there is an easy direction for the system to percolate, which is $\min(L_{x},L_{y})$. In a infinite system, the correlation length at the critical point will diverge symmetrically in the $\hat{x}$ and $\hat{y}$ directions as:
\begin{equation}
\xi \sim \left( \frac{\sigma-\sigma_{c}}{\sigma}\right)^{-\nu}.
\end{equation}
Note that we measure the distance to the critical point $\sigma_{c}$ with the same function as in Ref.\cite{spasojevicPRE2014}, instead of the standard choice $\left(\sigma-\sigma_{c}\right)/\sigma_{c}$. This issue was already proposed in the first works on the RFIM \cite{sethna93} and was extensively discussed in \cite{vives2004}. The direction for easy percolation is the one which first limits a 1d spanning avalanche. Thus, in the limits of $a\ll1$ and $a\gg1$, a finite system will experience pseudo-critical effects when the correlation length approaches to:
\begin{equation}
\xi \sim  L= \min\left( L_{x},L_{y}\right).
\end{equation} 
The function $\min\left( L_{x},L_{y}\right)$ exhibits a discontinuous derivative at $a=1$. In order to avoid it, we propose to measure the limit of the correlation length as:
\begin{equation}
L=L_{x} \frac{a}{\left(1+a^{1/\nu_{c}}\right)^{\nu_{c}}},
\end{equation}
and the corresponding scaling variable:
\begin{equation}
z=L_{x}^{1/\nu_{c}}\left(\frac{a^{1/\nu_{c}}}{1+a^{1/\nu_{c}}}\right)\frac{\sigma-\sigma_{c}}{\sigma}
\label{eq:variableescala}
\end{equation}
Note that this choice results in $z=\min(L_{x},L_{y})^{1/\nu_{c}}(\sigma-\sigma_{c}/\sigma)$ for the limits $a\ll1$ and $a\gg1$. As shown in Fig.~\ref{fig:escalatprevi}, curve collapses for 1d spanning avalanches in the $\hat{x}$ or $\hat{y}$ direction are achieved independently for each aspect ratio. Moreover the  peaks are alligned at the same abscissa:
\begin{equation}
N_{1}^{x,PBC}(\sigma,L_{x},a)=L_{x}^{\theta_{c}}\mathcal{P}_{a}^{x}\left(z\right),
\label{eq:primerescalatx}
\end{equation}
\begin{equation}
N_{1}^{y,PBC}(\sigma,L_{x},a)=L_{x}^{\theta_{c}}\mathcal{P}_{a}^{y}\left(z\right),
\label{eq:primerescalaty}
\end{equation}
with the same parameters $\theta_{c}$,$\nu_{c}$ and $\sigma_{c}$ as in Eq.~(\ref{eq:spasoscaling}). 
Strictly speaking, we only show data for $N_{1}^{y,PBC}$ but one must take into acount that the number of 1d spanning avalanches satisfies the following symmetry: 
\begin{equation}
N_{1}^{y,PBC}(\sigma,L_{x},a)=N_{1}^{x,PBC}\left(\sigma,aL_{x},\frac{1}{a}\right).
\label{eq:simetria}
\end{equation}
A similar expression was proposed in \cite{langlands} for the spanning probability in percolation theory.
 If one applies this symmetry to Eq.~(\ref{eq:primerescalaty}), the following relation between FSS functions must be fulfilled:
\begin{equation}
\mathcal{P}_{a}^{x}(z)=a^{\theta_{c}}\mathcal{P}^{y}_{1/a}(z),
\end{equation}
In order to assert the dependence of the scaling functions on the aspect ratio $a$, we propose that these functions  $\mathcal{P}^{x}_{a}(z)$ and $\mathcal{P}^{y}_{a}(z)$ can be written in the following way:
\begin{equation}
\mathcal{P}^{x}_{a}(z)=\lambda^{x}(a)\hat{\mathcal{P}}^{x}\left( z\right),
\label{eq:descomposiciox}
\end{equation}
\begin{equation}
\mathcal{P}^{y}_{a}(z)=\lambda^{y}(a)\hat{\mathcal{P}}^{y}\left(  z\right),
\label{eq:descomposicioy}
\end{equation}
where $\lambda^{x}(a)$, $\lambda^{y}(a)$  are functions of the parameter $a$. Note that the functions $\lambda^{x}(a)$ and  $\lambda^{y}(a)$ are different if one is considering 1d spanning avalanches in the $\hat{x}$ or $\hat{y}$ direction.

\begin{figure}[h!]
\resizebox{\columnwidth}{!}{\includegraphics{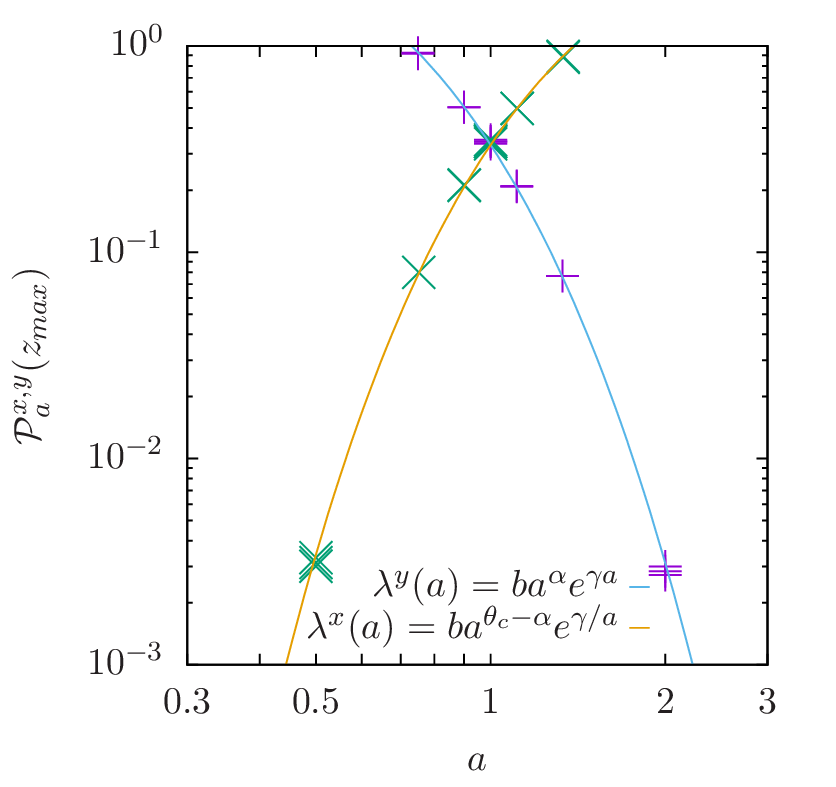}}
\caption{\label{fig:fits} Log-log plot of the peak heigths $\mathcal{P}^{x,y}_{a}(z_{max})$ as a function of the aspect ratio $a$ and fits of functions $\lambda^{x}(a)$ and $\lambda^{y}(a)$ as described in the text.}
\end{figure}

\begin{figure}[h!]
\resizebox{\columnwidth}{!}{\includegraphics{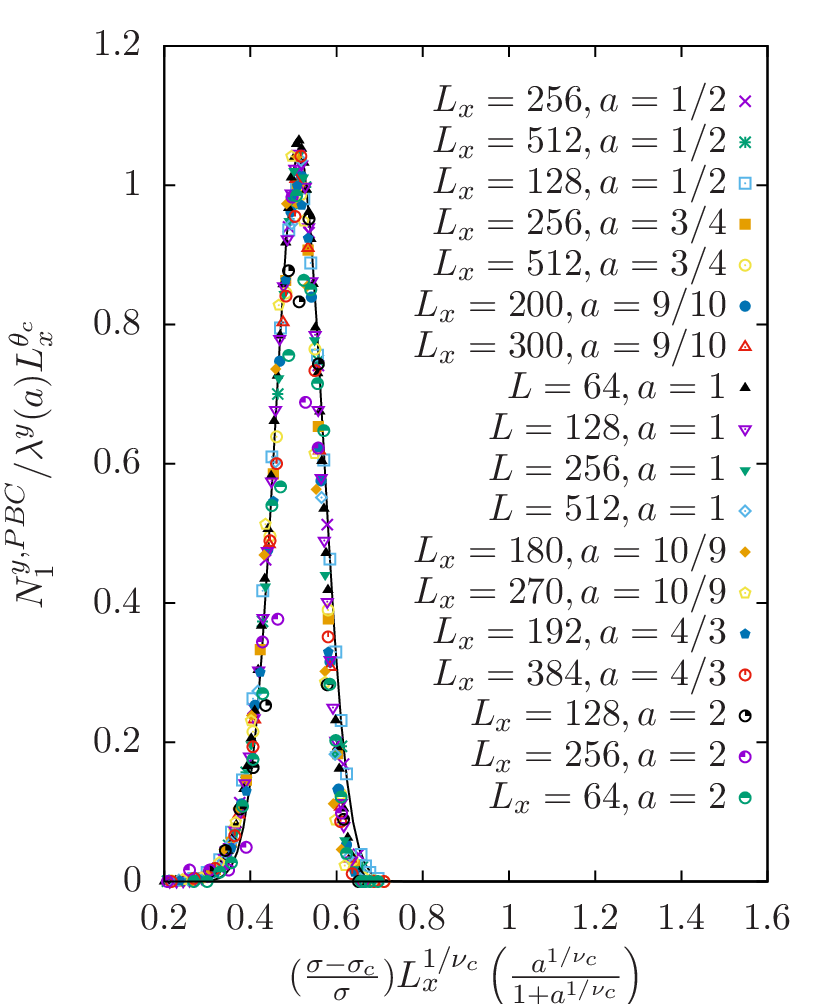}}
\caption{\label{fig:escalat}Curve collapses for 1d spanning avalanches in the $\hat{y}$ direction for different aspect ratios and sizes. Black solid line corresponds to the Gaussian fit of Fig.~\ref{fig:3} (b).}
\end{figure}

From the study of the peak heights $\mathcal{P}^{x,y}_{a}(z_{max})$ a good fit (See Fig.~\ref{fig:fits}) can be obtained with:
\begin{align}
\lambda^{x}(a)&=ba^{\theta_{c}-\alpha}e^{\gamma/a}\\
\lambda^{y}(a)&=ba^{\alpha}e^{\gamma a},
\label{eq:caigudaexponencial}
\end{align}
where $b=80\pm42$, $\alpha=1.2\pm0.4$ and $\gamma=-5.5\pm0.5$. One must take into account that $b$, $\alpha$ and $\gamma$ have an empirical character and play no role in the following developments.
Note that this choice satisfies the symmetry:
\begin{equation}
\frac{\lambda^{x}(a)}{\lambda^{y}(1/a)}=a^{\theta_{c}}
\end{equation} 
From this last equality and equations ~(\ref{eq:descomposiciox}) and ~(\ref{eq:descomposicioy}) one obtains:
\begin{equation}
\hat{\mathcal{P}}^{x}(z) = \hat{\mathcal{P}}^{y}(z) \equiv \hat{\mathcal{P}}(z),
\label{eq:igualtat}
\end{equation} 
which means that, in fact, we can formulate a scaling function $\hat{\mathcal{P}}$ independent of the spatial direction.
Taking into account all these premises, satisfactory collapses of the whole set of numerical data are obtained with the following FSS hypothesis:
\begin{equation}
N_{1}^{x}(\sigma,L_{x},a)= L_{x}^{\theta_{c}}\lambda^{x}(a)\hat{\mathcal{P}}\left(z\right) 
\label{eq:escalatuniversalx}
\end{equation}
\begin{equation}
N_{1}^{y}(\sigma,L_{x},a)= L_{x}^{\theta_{c}}\lambda^{y}(a)\hat{\mathcal{P}}\left(z\right) 
\label{eq:escalatuniversaly}
\end{equation}
Overlaps for peaks corresponding to $N_{1}^{y,PBC}$ are shown in Fig.~\ref{fig:escalat}.
\subsection{Fixed Boundary Conditions (FBC)}
\label{FBC}
\begin{figure}[h!]
\resizebox{\columnwidth}{!}{\includegraphics{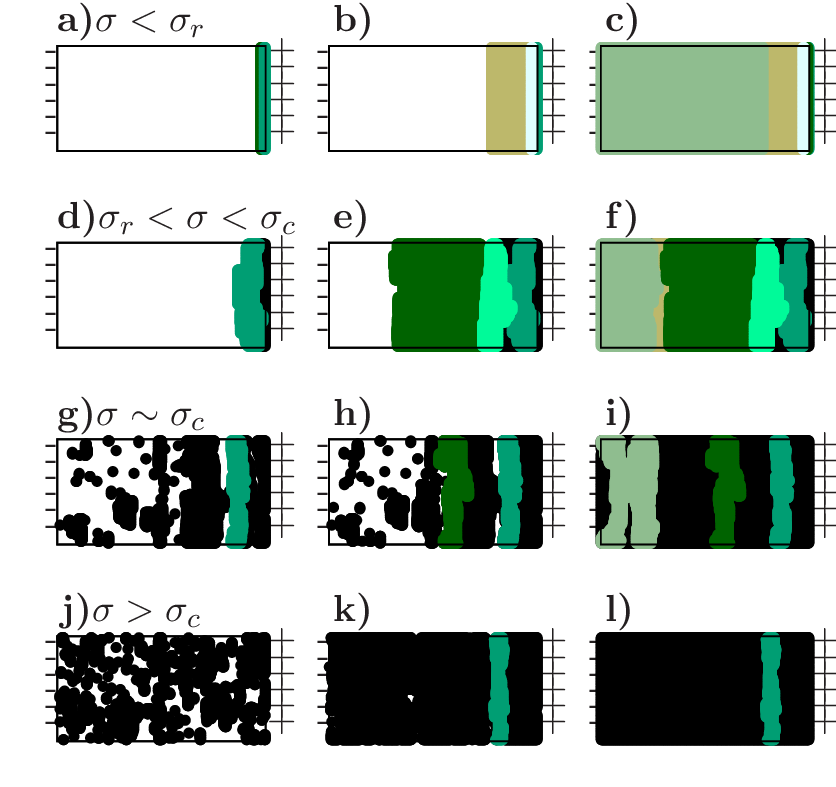}}
\caption{\label{fig:sequencia}Sequence of configurations during the magnetization process for the field driven, athermal RFIM with metastable dynamics under FBC for a system with $L_{x}=512$ and aspect ratio $a=1/8$. The values of the disorder are $\sigma=0.25,0.50,0.90,1.20$ for sequences (a-b-c),(d-e-f),(g-h-i) and (j-k-l) respectively. External field $H$ is increased from left to right. Symbols (+) and (-) correspond to spins which have values $s_{i}=+1$ and $s_{i}=-1$ respectively and conform the fixed boundaries. Black coloured regions correspond to non-spanning avalanches whereas 1-d spanning avalanches are represented by different tonalities of green. A front that advances from right to left is easily identified in sequences (a-b-c),(d-e-f) and (g-h-i).}
\end{figure}

The idea behind imposing FBC is to force simulations to exhibit a domain wall which can be identified as an advancing front (See Fig.~\ref{fig:sequencia}). From a physical point of view, this would represent a situation in which the system under study is a subset of a bigger one with an already formed interface. By keeping periodic boundaries in the vertical $\hat{y}$ direction, the horizontal $\hat{x}$ direction is subjected to the following condition: at the boundary $x=L_{x}+1$, there is a column of spins $\lbrace s_{i}=+1 \rbrace$ (+ symbols in Fig.~\ref{fig:sequencia}) whereas at $x=0$ there is a column of spins $\lbrace s_{i}=-1 \rbrace$ (- symbols in Fig.~\ref{fig:sequencia}). Under these specific conditions, 1d spanning avalanches in the $\hat{y}$ direction are more common in a regime of low disorder. A minuscule fraction of 1d spanning avalanches in the $\hat{x}$ direction and 2-d spanning avalanches are found but with no relevant statistical weight.

When the system presents a stripe geometry ($a<1$), it is easy to distinguish four different regimes for the dynamics of 1d spanning avalanches as a function of the disorder $\sigma$. Below a certain value of the quenched disorder, $\sigma<\sigma_{r}$, a sequence of massive spanning avalanches with flat interface appears during the magnetization process (See Fig.~\ref{fig:sequencia} sequence a-b-c). As disorder increases, there exists a regime $\left(\sigma_{r}<\sigma < \sigma_{c}\right)$ where the advancing front presents a rough profile and there is a negligible presence of nucleated domains in front of the advancing interface (Fig.~\ref{fig:sequencia} d-e-f)). When the disorder approaches its critical value ($\sigma \sim \sigma_{c}$), the  critical interface advances interacting with many nucleated domains (Fig.~\ref{fig:sequencia} g-h-i)). Above $\sigma_{c}$, it is unlikely to find a 1d spanning avalanche so the front is ill-defined except for some very rare cases (Fig.~\ref{fig:sequencia} j-k-l)). For higher disorders, a pure nucleation and growth process is recovered.

The main goal of this section is to study the consequence of changing the nature of the boundaries on the FSS functions and to relate them in the regime dominated by critical effects with those found with PBC for any aspect ratio $a$.

\begin{figure}[h!]
\resizebox{\columnwidth}{!}{\includegraphics{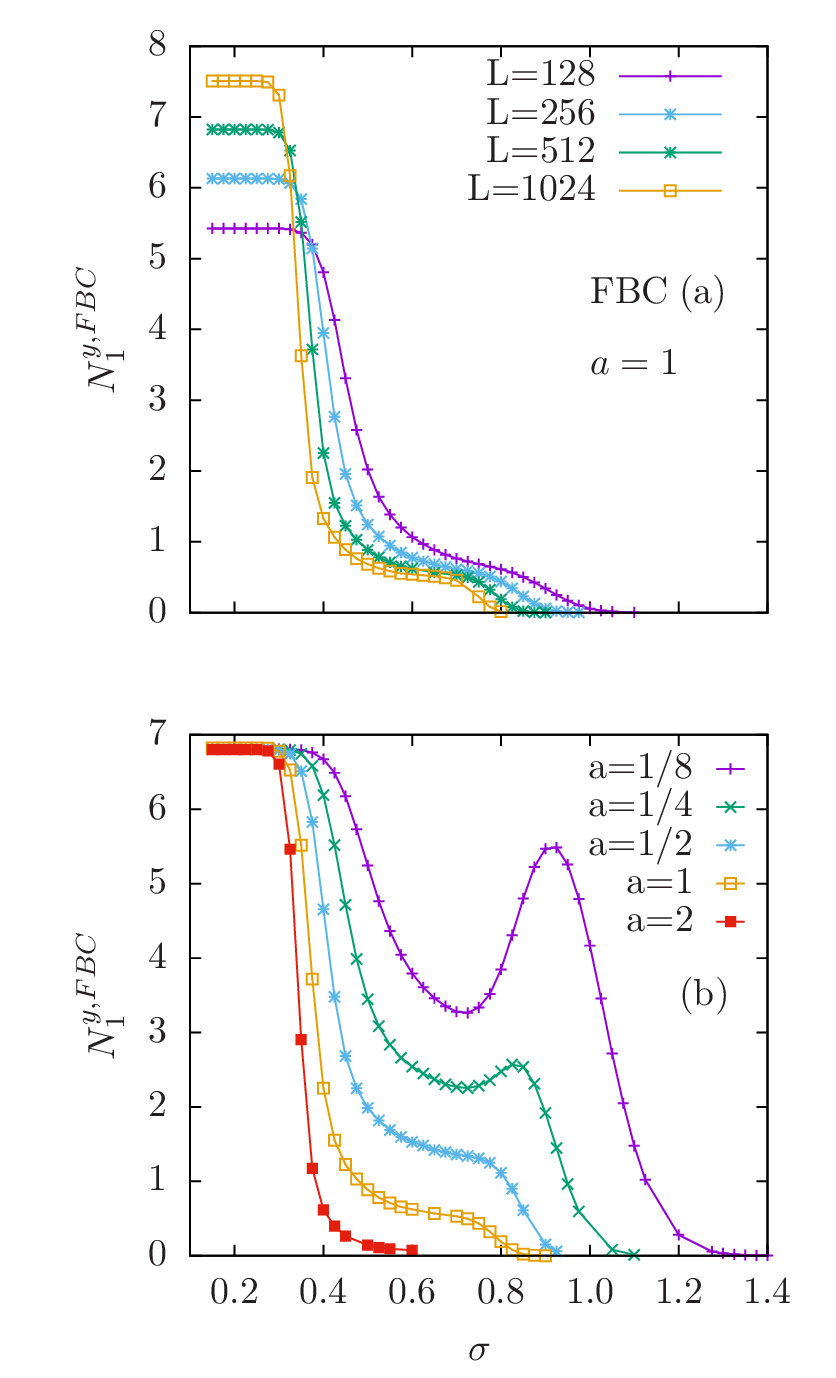}}
\caption{\label{fig:512}Panel in (a) shows the average number of 1d spanning avalanches in the $\hat{y}$ direction as a function of the disorder for a system with $a=1$ and different sizes. Panel in (b) shows the average number of 1d spanning avalanches in the $\hat{y}$ direction as a function of disorder for different aspect ratios for a system with $L_{x}=512$. Lines are guides to the eye.}
\end{figure}

As it can be observed in Fig.~\ref{fig:512} (a), two steps are clearly distinguished in the behavior of the number of 1d spanning avalanches $N_{1}^{y,FBC}(\sigma)$. The higher step on the left is related to the morphological transition of the propagating front. This transition separates the regimes where there is a sequence of 1d spanning avalanches with a flat profile (faceted growth) and the regime where 1d spanning avalanches exhibit a certain rough profile. The second step occurs in the region  $\sigma \sim 0.8$ where, in the case of PBC, the peaks are found. This suggests that steps are related to the critical transition.

Note that at low disorder the average number of spanning avalanches $N_{1}^{y,FBC}$ reaches values above $5$ whereas for similar system sizes and disorder $N_{1}^{PBC}$ with PBC is negligible (See Fig.~\ref{fig:3} (a)). A qualitative argument to justify such a difference is given as follows: in the low disorder regime, local fields take values around zero. As spins near the boundary $x=L_{x}$ have a neighbour which is pointing up $\lbrace s_{i}=+1 \rbrace$ the external field needs to be around $H=2$ in order to flip them and create a nucleation centre for a 1d spanning avalanche that propagates in the $\hat{y}$ direction. In the model with PBC, the external field needs to be around $H=4$ in order to flip a spin and generate a nucleation centre which can span the system in one or two directions. When the system is subjected to FBC, it is very difficult to find a horizontal spanning avalanche which connects both sides of the system with fixed boundaries as well as 2d spanning avalanches.

In order to elucidate how the height of the observed steps depend on the system size and shape, simulations at different $L_{x}$ and $L_{y}$ have been performed. Results for fixed $L_{x}$ and different aspect ratios are shown in Fig.~\ref{fig:512} (b). Note that the height of the step on the left does not depend on $a$. This indicates that the horizontal length $L_{x}$ controls the number of 1d spanning avalanches at low disorders.  By changing the aspect ratio, the morphology of the curves in Fig.~\ref{fig:512} (b) changes. When $a$ is lower than unity, which means that the system is dominated by periodic boundaries, critical effects are strengthened and what seemed to be a step for $a=1$ becomes a clear peak. As the aspect ratio increases, the sides of the system which are subjected to fixed boundaries take over and critical effects are gradually hindered. This suggests that, in the critical region, the invariance under 90 degree rotations is recovered for disorders close to $\sigma_{c}$ and the same analysis as in the previous section can be performed.

\begin{figure}[h!]
\centering
\resizebox{\columnwidth}{!}{\includegraphics{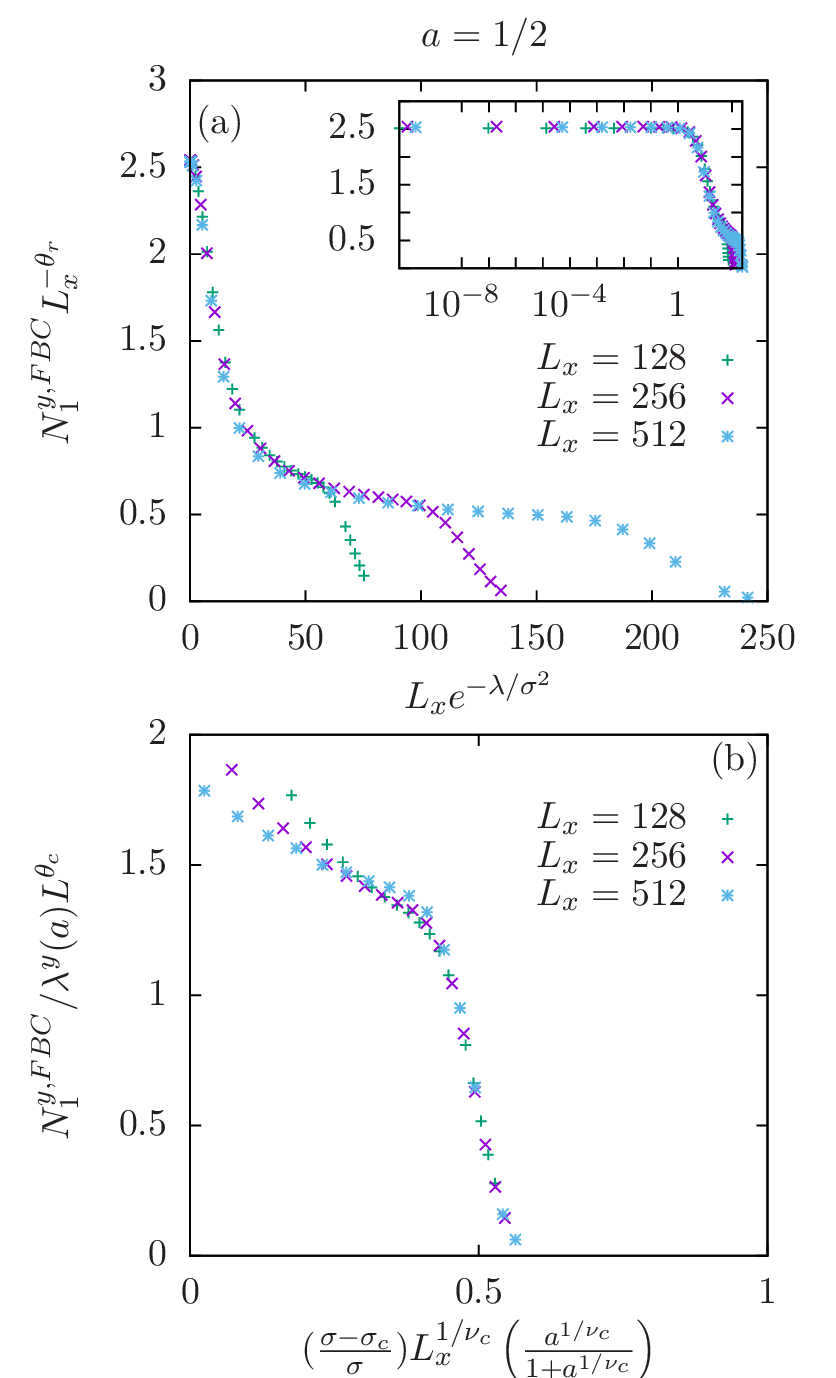}}
\caption{\label{fig:a12}Curve collapse for average number of 1d spanning avalanches per magnetization process for different system sizes in a system with $a=1/2$. Partial curve collapse in the regime dominated by roughening is presented in (a). The step can be appreciated in the inset plot where the same figure is represented with logaritmic scale in the horizontal axis. Partial curve collapse for the regime dominated by the bulk critical transition is shown in (b).}
\end{figure}

\begin{figure}[h!]
\centering
\resizebox{\columnwidth}{!}{\includegraphics{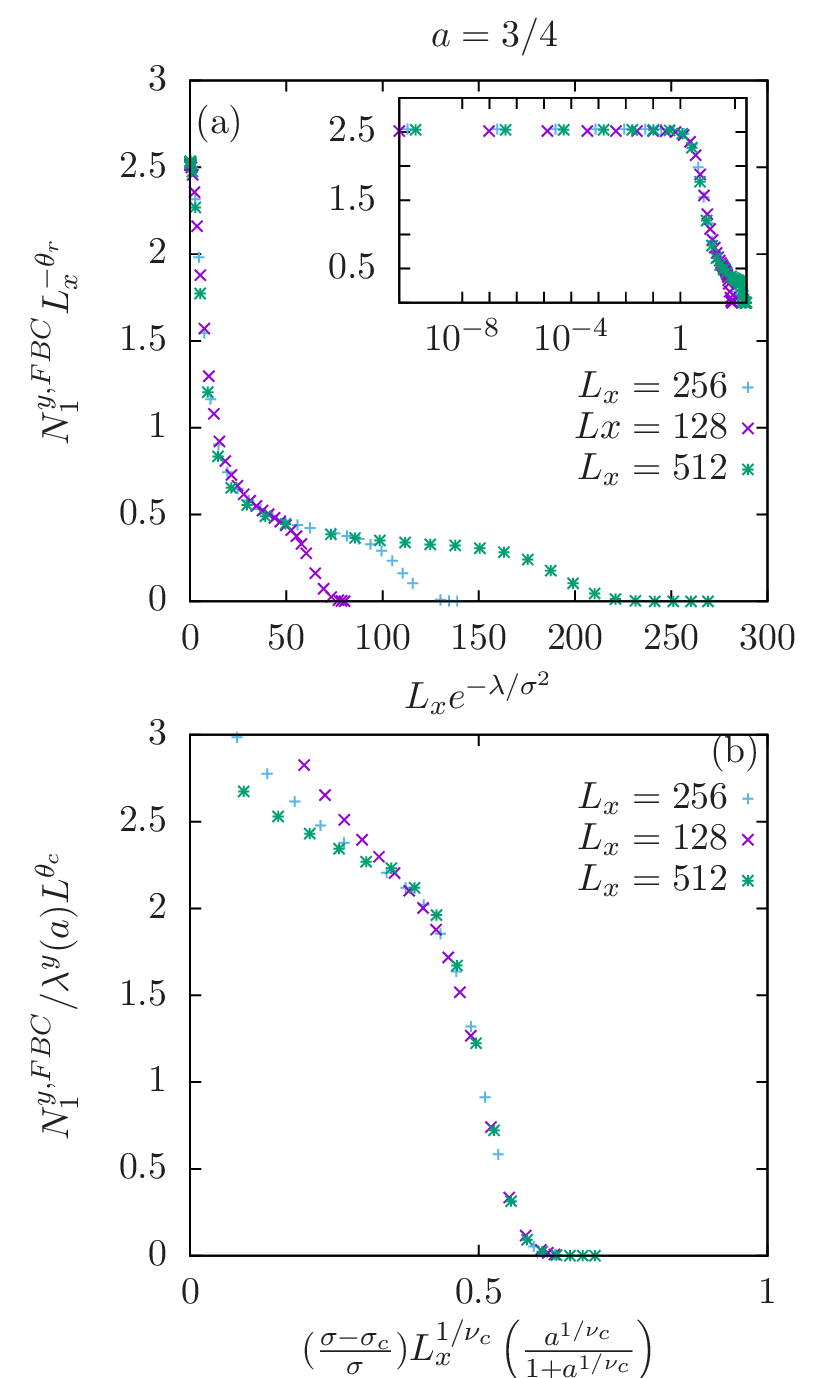}}
\caption{\label{fig:a34}Curve collapse for average number of 1d spanning avalanches per magnetization process for different system sizes in a system with $a=3/4$. Partial curve collapse in the regime dominated by roughening is presented in (a). The step can be appreciated in the inset plot where the same figure is represented with logaritmic scale in the horizontal axis. Partial curve collapse for the regime dominated by the critical transition is shown in (b).}
\end{figure}

\begin{figure}[h!]
\resizebox{\columnwidth}{!}{\includegraphics{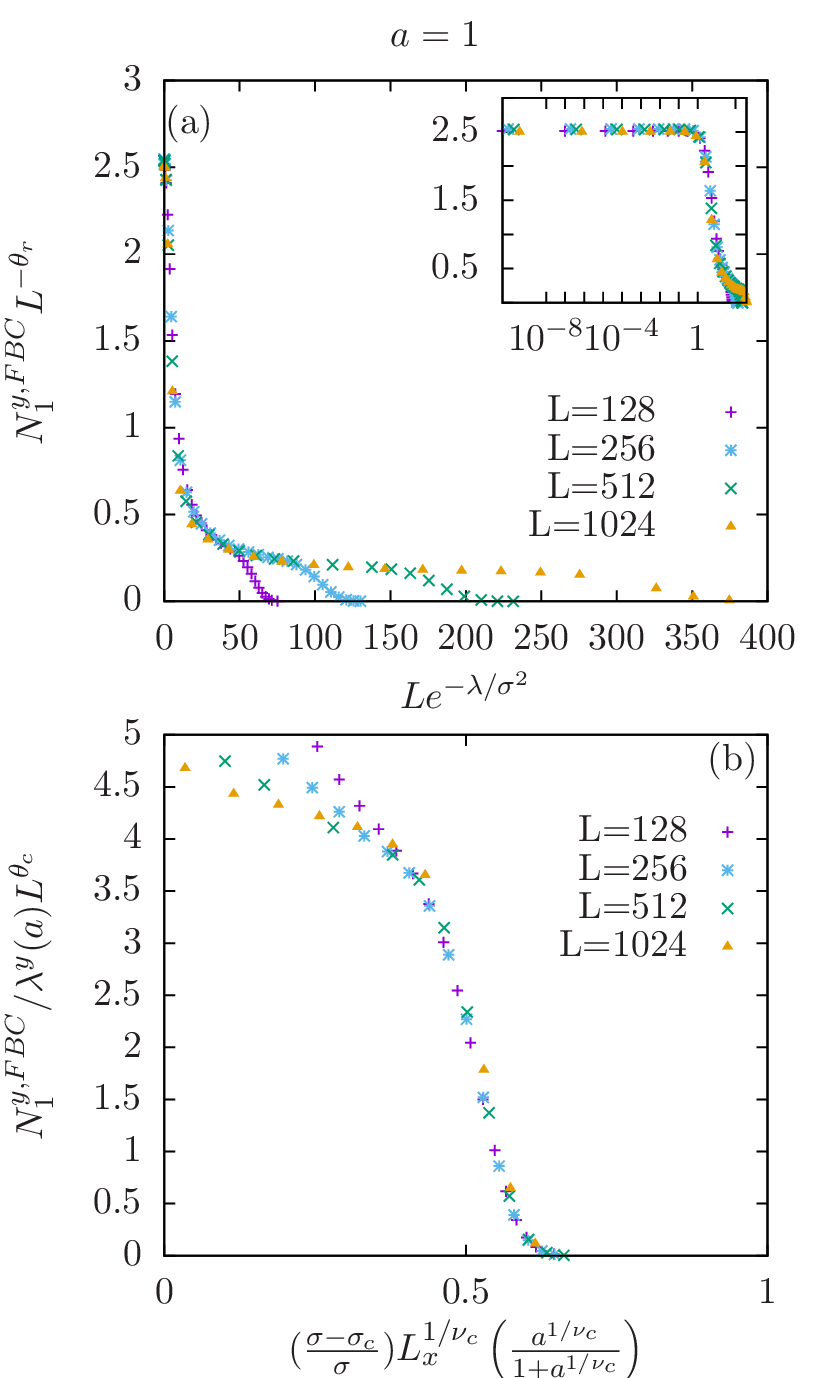}}
\caption{\label{fig:combinedfss}Curve collapse for Average number of 1d spanning avalanches per magnetization process for different system sizes in a square system $L_{x}=L_{y}\equiv L$. Partial curve collapse in the regime dominated by roughening effects is presented in (a). The step can be appreciated in the inset plot where the same figure is represented with logaritmic scale in the horizontal axis. Partial curve collapse for the regime dominated by the critical transition is shown in (b).}
\end{figure}

\begin{figure}[h!]
\centering
\resizebox{\columnwidth}{!}{\includegraphics{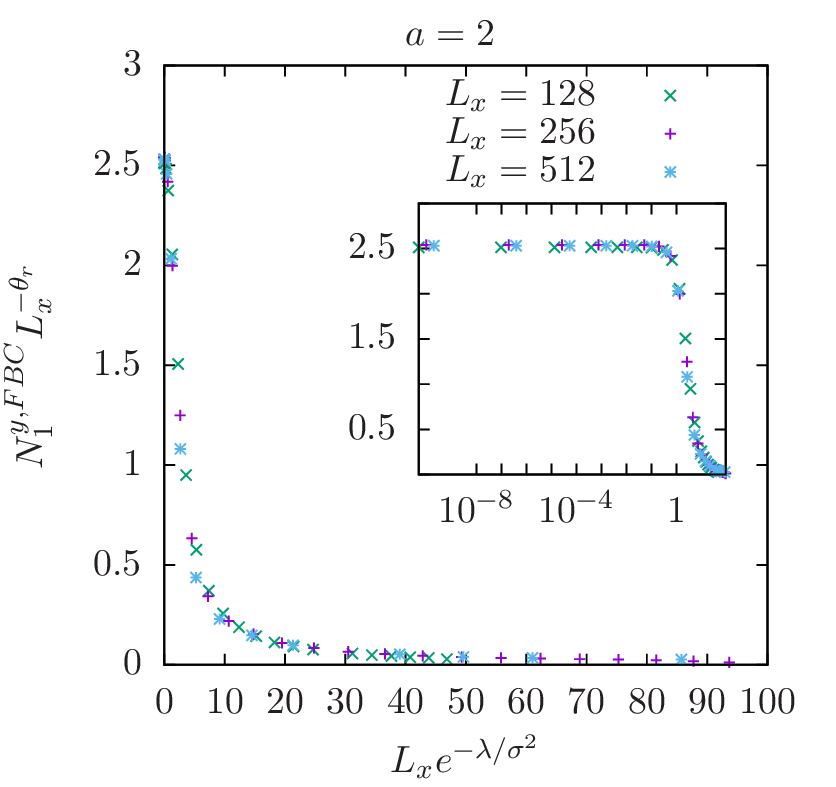}}
\caption{ \label{fig:a2}Curve collapse for average number of 1d spanning avalanches per magnetization process for different system sizes in a system with $a=2$. The step can be appreciated in the inset plot where the same figure is represented with logaritmic scale in the horizontal axis. In this case, critical effects have been hindered and the contribution of spanning avalanches in the critical region is negligible.}
\end{figure}

The discussion above suggests that we shall propose a combined FSS relation with parameters related to both transitions ($\sigma_{c}$,$\sigma_{r}$) in order to achieve partial curve collapse. For a certain aspect ratio, the scaling hypothesis is:
\begin{equation}
N_{1}^{y}(\sigma,L_{x},aL_{x})=L_{x}^{\theta_{r}}\hat{\hat{\mathcal{F}}}_{a}\left( \tilde{z} \right)+L_{x}^{\theta_{c}}\lambda^{y}(a) \hat{\mathcal{F}}\left(z \right),
\label{eq:combined}
\end{equation}
where $\tilde{z}=L_{x} e^{-\lambda/\sigma^{2}}$ is the scaling variable in the roughening regime with $\lambda=0.64 \pm 0.03$ and $\theta_{r}=0.16\pm 0.01$. This choice of the scaling variable $\tilde{z}$ was already proposed in \cite{condmat96} and \cite{braymoore}. It means that the roughening transition occurs for a certain value of the disorder only due to finite size effects. Faceted growth will not be present in the thermodynamic limit for Gaussian random fields as also explained by Ji and Robbins \cite{jiRobbins91}. We will use the calligraphic letter $\mathcal{F}$ for the FSS functions corresponding to FBC. Partial collapses in the roughening regime are shown in  Fig.~\ref{fig:a12} (a) for $a=1/2$, Fig.~\ref{fig:a34} (a) for $a=3/4$, Fig.~\ref{fig:combinedfss} (a) for $a=1$, and Fig.~\ref{fig:a2} for $a=2$. Curves clearly overlap when the scaling variable is below $50$ aproximately.
The second term of Eq.~(\ref{eq:combined}) has the same form and the same scaling variable as in Eq.~(\ref{eq:escalatuniversaly}) from the previous section. 
Partial curve collapses in the critical region are shown in Fig.~\ref{fig:a12} (b) for $a=1/2$, in Fig.~\ref{fig:a34} (b) for $a=3/4$ and in Fig.~\ref{fig:combinedfss} (b) for $a=1$ . Note that the hypothesis of combined FSS remains valid for different aspect ratios where the critical effects have more presence (Fig.~\ref{fig:a12}) or have practically disappeared (Fig.~\ref{fig:a2}).

\begin{figure*}[!]
\centering
\includegraphics{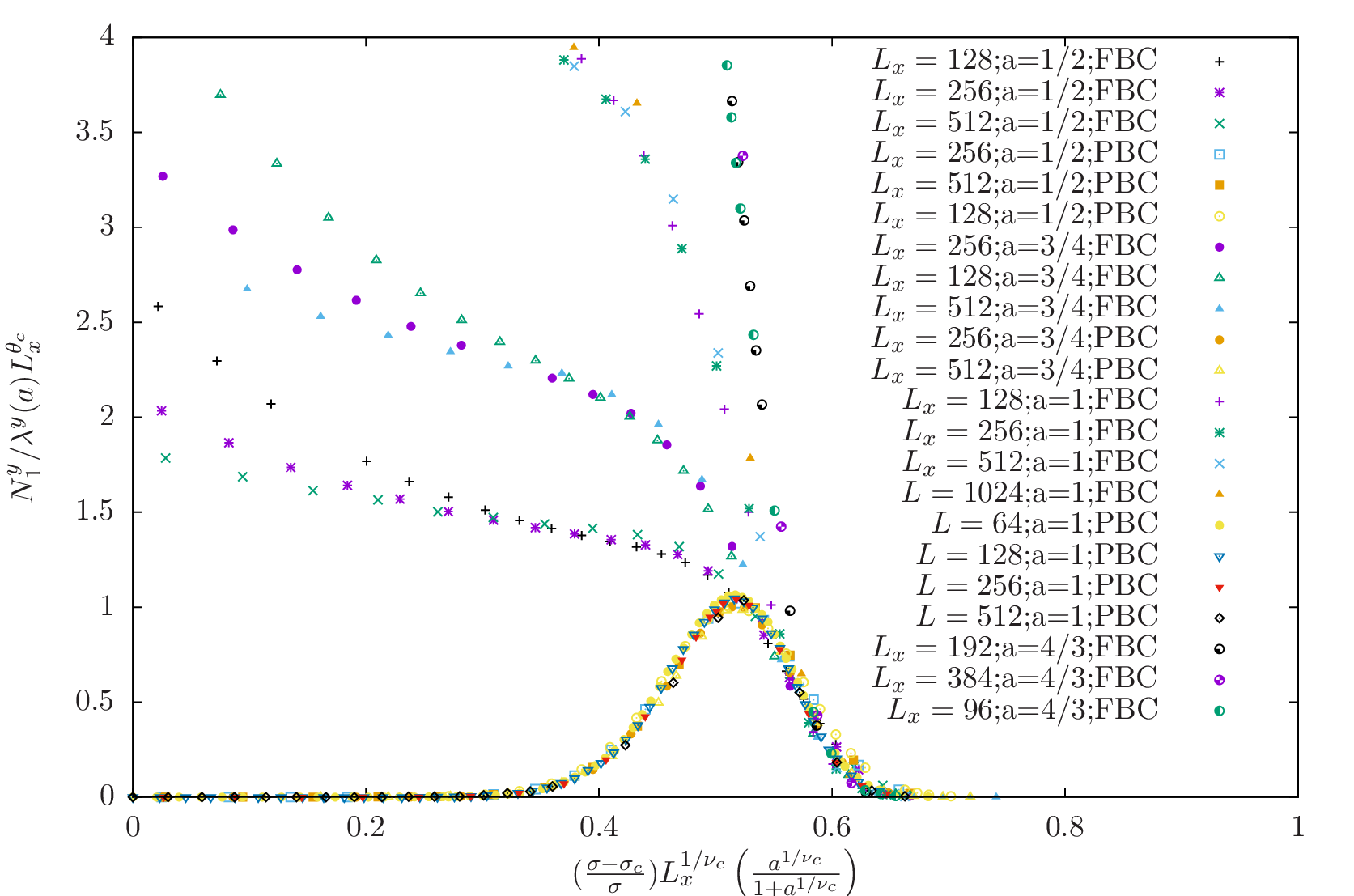}
\caption{\label{fig:equival}Overlaps of FSS functions with FBC and PBC. Eq.~(\ref{eq:y}) are satisfied for $a=\lbrace 1/2$, $3/4$, $1$,$4/3 \rbrace$  }
\end{figure*}
Finally, let us test the relation between FSS functions for PBC and FBC in the critical region. 
Fig.~\ref{fig:equival} shows that indeed, both FSS functions are equivalent:
\begin{equation}
\hat{\mathcal{P}}(z)=\hat{{\mathcal{F}}}(z)\equiv \hat{\mathcal{Q}}(z)
\label{eq:y}
\end{equation}
Note that we achieve partial overlaps of data corresponding to FBC and PBC and corresponding to $a=\lbrace 1/2$ ,$3/4$, $1$, $4/3 \rbrace$. Consequently, the FSS function $\hat{\mathcal{Q}}(z)$ does not depend on $a$ neither on the nature of the boundary conditions. Note that, the smaller (larger) the aspect ratio the broader (tiny) the range of agreement between FSS functions. This makes sense since, as explained in Fig.~\ref{fig:512}(b), for aspect ratios greater than unity critical effects are hindered.
The left tails of peaks found for PBC can not be related with the curves for FBC. In that regime and for the studied system sizes the presence of the FBC is still too strong to appreciate critical effects.

\section{Summary and conclusions}
\label{Conclusions}
In this work we have presented results corresponding to the field driven, athermal RFIM with local relaxation dynamics with PBC and FBC and rectangular geometries with different aspect ratios $a$.  We have proposed a new scaling variable 
\begin{equation}
z=L_{x}^{1/\nu_{c}}\left(\frac{a^{1/\nu_{c}}}{1+a^{1/\nu_{c}}}\right)\frac{\sigma-\sigma_{c}}{\sigma}
\end{equation}
With this choice, the average number of 1d spanning avalanches in $\hat{x}$ and $\hat{y}$ directions can be scaled as:
\begin{equation}
N_{1}^{x,y}(\sigma,L_{x},a)=L_{x}^{\theta_{c}}\lambda^{x,y}(a)\tilde{\mathcal{Q}}(z)
\end{equation} 
where the functions $\lambda^{x,y}$ depend on wether we study 1d spanning avalanches in the $\hat{x}$ or $\hat{y}$ direction and satisfy:
\begin{equation}
\frac{\lambda^{x}(a)}{\lambda^{y}(1/a)}=a^{\theta_{c}}.
\end{equation}
The physical meaning of these prefactors $\lambda^{x,y}$ account for the relative increase or decrease of the number of 1d spanning avalanches when changing the aspect ratio. With these definitions, the scaling function $\hat{\mathcal{Q}}$ is thus independent of the aspect ratio $a$ and, even more, independent on the nature of the boundary conditions. To observe the collapses corresponding to Eq.(\ref{eq:escalatuniversaly}) for the case of FBC, it is only possible for large enough values of $z$ so that the effects caused by the faceted growth of the interfaces become irrelevant. 

In future works it would be interesting: a) to perform the same study for the 3d case and b) to study the morphological properties of the advancing front and its dynamics for different amounts of disorder.

\begin{acknowledgments}
This work was completed with computational resources provided by IBERGRID collaboration. Financial support was received from Ministerio de Econom\'ia y Competitividad (Spain) (Project No MAT 2015-69-777-REDT and Project No MAT2013-40590-P) and from Fundaci\'o la Caixa. We thank the referee for suggesting the scaling variable in Eq.\ref{eq:variableescala} that improved our previous choice. We also acknowledge D. Spasojevi\'c for fruitful discussions during a visit to Universitat de Barcelona.
\end{acknowledgments}

\end{document}